\documentclass[preprint]{aastex}
\usepackage{natbib}

\shorttitle{Nitrogen Enriched Quasars in SDSS DR1}
\shortauthors{Bentz, Hall \& Osmer}

\begin{document}
\title{Nitrogen Enriched Quasars in the Sloan Digital Sky Survey First 
Data Release}
\author{Misty C. Bentz}
\affil{Department of Astronomy, The Ohio State University}
\affil{140 W. 18th Ave, Columbus, OH 43210-1173}
\email{bentz@astronomy.ohio-state.edu}
\author{Patrick B. Hall}
\affil{Princeton University Observatory}
\affil{Princeton, NJ 08544-1001}
\email{pathall@astro.princeton.edu}
\author{Patrick S. Osmer}
\affil{Department of Astronomy, The Ohio State University}
\affil{140 W. 18th Ave, Columbus, OH 43210-1173}
\email{posmer@astronomy.ohio-state.edu}

\begin{abstract}
The quasar Q0353-383 has long been known to have extremely strong
nitrogen intercombination lines at $\lambda$1486 and $\lambda$1750 \AA,
implying an anomalously high nitrogen abundance of $\sim 15$ times
solar.  A search for similar nitrogen-rich quasars in the Sloan Digital
Sky Survey First Data Release (SDSS DR1) catalog has yielded 20
candidates, including four with nitrogen emission as strong or stronger
than that seen in Q0353-383.  Our results indicate that only about 1 in
1700 of quasars have nitrogen abundances similar to Q0353-383, while up
to 1 in 130 may be in the process of extreme nitrogen enrichment.
\end{abstract}


\section{Introduction}
The quasar Q0353-383 \citep{OsmerSmith1980} is an unusual object, with
prominent \ion{N}{3}], \ion{N}{4}] and \ion{N}{5} emission lines and
abnormally weak \ion{C}{3}] and \ion{C}{4} lines compared to other
quasars. To illustrate this point, Figure 1a displays the spectrum of
Q0353-383 (Baldwin, 1992, private communication) in comparison to a
``normal'' quasar spectrum given by the Sloan Digital Sky Survey
\citep{York2000} composite in Figure 1b \citep{VandenBerk2001}.
\citet{Osmer1980} concluded that Q0353-383 has an anomalously high
nitrogen abundance due to recent CNO processing in
stars. \citet{Baldwin2003} used improved data and models to confirm and
refine these conclusions, and to determine that Q0353-383 has a
metallicity of at least 5 times solar, and more likely 15 times solar.
Simulations by \citet{Hamann1999} show that this level of overabundance
is expected to occur near the end of an era of rapid metal enrichment
which can result in metallicities of as much as 10-20 times solar.  The
scarcity of objects like Q0353-383 may be an indication of the amount of
time a quasar spends in this state of extreme metal enrichment before
the gas supply is exhausted and the quasar becomes inactive.

As the only object of its kind known, Q0353-383 raises some obvious
questions: what percentage of the quasar population is nitrogen strong,
and what are the global properties of the nitrogen-strong quasar
population?  Until recently, anwering these questions was difficult due
to the relatively nonstandard collection of quasar data in various
wavelength regimes and the lack of spectra with high S/N
(signal-to-noise ratio).  The advent of the Sloan Digital Sky Survey
(SDSS) has remedied this situation by working to compile, in one
database, approximately 100,000 high-quality quasar spectra as it scans
10,000 deg$^2$ of the north Galactic cap \citep{York2000}.
\citet{Bentz2003} searched the SDSS Early Data Release (EDR, 
\citealt{Stoughton2002}) for objects similar to Q0353-383 and
determined that although several objects have nitrogen emission that is
unusually strong, none of the objects in the EDR Quasar Catalog
\citep{Schneider2002} with $1.8 < z < 4.1$ have emission from both
\ion{N}{4}] $\lambda$1486 and \ion{N}{3}] $\lambda$1750 with strengths
that are comparable to Q0353-383.  In this paper, we have searched for
nitrogen-rich quasars in the Quasar Catalog \citep{Schneider2003} from
the First Data Release (DR1, \citealt{Abazajian2003}), which covers
almost three times the area on the sky as the EDR, and has over four
times as many quasars.  We present numerous objects with stronger
nitrogen emission than is usually seen in quasars, including four
objects that have emission as strong or stronger than that seen in
Q0353-383.

\section{Spectral Analysis}
The SDSS DR1 Quasar Catalog \citep{Schneider2003} covers an area of
$\sim$ 1360 $\rm{deg^2}$ of the sky, containing 16,713 objects with
luminosities greater than $\rm{M_i} = -22$ mag (for a cosmology with
$H_{0} = 70$, $\Omega_M = 0.3$, and $\Omega_{\Lambda} = 0.7$), at least
one emission line with a FWHM larger than 1000 km $\rm{s^{-1}}$, and
reliable redshifts.  The entire area scanned for the EDR is contained
within DR1, and the quasar spectra from the EDR Quasar Catalog were run
through the spectroscopic pipeline again after several modifications and
improvements were made.  In order to find quasars similar to Q0353-383,
we focused our search on objects that exhibited the rest-frame
ultraviolet nitrogen intercombination lines at $\lambda 1486$ and
$\lambda 1750$.  It is important to note that \ion{N}{4}] $\lambda 1486$
and \ion{N}{3}] $\lambda 1750$ are collisionally de-excited at densities
greater than $\simeq 10^{11}\ \rm cm^{-3}$ \citep{Ferland1999}, so it is
possible for nitrogen-rich quasars to exist where a strong \ion{N}{5}
line would be the only indication \citep{Hamann1999proc}.  However, the
detection of objects similar to Q0353-383 would help to set a lower
limit on the population of nitrogen-rich quasars based on a larger
sample size than previous studies, and further study of such objects
could help to increase our knowledge of the end stages of quasar
activity.

The SDSS DR1 database was searched for all quasars within the redshift
range $1.6 < z < 4.1$.  This range ensures that both \ion{N}{4}]
$\lambda$1486 and \ion{N}{3}] $\lambda$1750 will be in the 3800-9200\AA\
range observed by the SDSS spectrograph.  A total of 6650 objects met
the redshift criterion.  We used the redshift values determined by
\citet{Schneider2003} to correct for cosmological expansion and place
the spectra in a common rest frame.  As many of the objects in the SDSS
are faint and the features we searched for are weak, we first estimated
the signal-to-noise (S/N) per pixel (where 3 pixels $\approx$ 1 \AA) of
the spectra using the continuum between 1675 and 1725\AA.  We threw out
the noisiest spectra by making a cut at S/N $> 3.45$, reducing the
sample to $\sim 5600$ quasars.  We then ran a cross-correlation routine
to compare a 30\AA\ window centered on 1750\AA\ with the same window in
the rest frame spectrum of Q0353-383 to test for the presence or absence
of an emission features in the location of \ion{N}{3}]. \footnote{The
cross-correlation routine was also used in a 30 \AA\ window centered on
$\lambda 1486$ to search for emission from \ion{N}{4}].  However,
intrinsic absorption was a large cause of noise in the results of the
cross-correlation on this area of the spectra, and the results from the
\ion{N}{3}] window were deemed more robust.}  A further cut was made
with the twofold criteria of a relative velocity offset less than 850 km
s$^{-1}$ and a cross-correlation coefficient value of at least 0.35 for
any feature detected in the \ion{N}{3}] 30\AA\ window.  The $\sim 1350$
objects that passed the cuts were then visually inspected in order to
verify the presence of emission from nitrogen, as objects with features
such as absorption or noise spikes could also be included in the sample.
All objects with broad absorption-line profiles were immediately
discounted ($\sim 6\%$ of the sample).  Only 198 ($\sim 15\%$) of the
objects did not seem to have evidence for an emission feature near
$\lambda1750$\AA\ upon visual inspection.  Objects that appeared to have
prominent emission from both species of nitrogen were marked for further
inspection.

Finally, equivalent widths of the \ion{N}{4}] and \ion{N}{3}] emission
lines ($W_{N IV}$ and $W_{N III}$, respectively) were measured in the
rest-frame spectra using a simple summing function with a two-point
continuum interpolation.  These are only guideline measurements, and
therefore have a typical error of 0.5-1.0\AA.  Several of the objects
that were selected with the cross-correlation method and subsequently
passed the visual inspection were found to have relatively large values
of $W_{N IV}$ and $W_{N III}$.  Hereafter, we shall focus on those
objects with $W_{N IV}$ and $W_{N III}$ $\gtrsim 2.0$\AA\ (see Table 1
for their general properties and Table 2 for emission line
measurements).  These quasars will hereafter be referred to as
nitrogen-rich candidates, and their spectra are displayed in Figure 2.
Table 3 lists another group of quasars that we shall refer to as
nitrogen-salient quasars.  These objects have obvious emission from at
least one of the species of interest, but do not meet the above criteria
of $W_{NIV}$ and $W_{NIII} \gtrsim 2.0$\AA. In all, our search revealed
20 nitrogen-rich candidates and 31 nitrogen-salient quasars in the SDSS
DR1 database.

\subsection{Description of Candidates}
The sample of candidate nitrogen-rich quasars presented here contains 20
objects that have similar features.  Ten of the candidates have
relatively flat F$_{\lambda}$ continua, while the other 10 have slight
blueward slopes.  Most of the candidates are narrow-lined objects,
although the FWHM of the \ion{C}{4} emission line ranges from $\sim 950
- 4300$ km s$^{-1}$, with a median FWHM of $\sim 2100$ km s$^{-1}$.  All
of the spectra show evidence for \ion{He}{2} $\lambda$1640 and
\ion{O}{3}] $\lambda$1663 emission, which will be important diagnostics
for metallicity estimates, but in most of the candidates they are too
blended for preliminary measurements at the S/N afforded by the SDSS
spectra.

The profiles of the Ly$\alpha$ emission lines (for those candidates with
$z > 2.3$ where Ly$\alpha$ has redshifted into the spectrograph range)
are very narrow and sharp.  All of these quasars have visibly separated
\ion{N}{5} emission, and three of the seven (SDSS J0242-0038, SDSS
1059+6638, and SDSS J1550+0236) have \ion{N}{5} emission that is
relatively unblended with the Ly$\alpha$ emission.  The FWHM of the
Ly$\alpha$ line ranges from $\sim 1300-3000$ km s$^{-1}$ with a median
value of 1700 km s$^{-1}$, compared to $\sim 6100$ km s$^{-1}$ for the
SDSS quasar composite of \citet{VandenBerk2001}. \citet{Osmer1980}
noticed that the value of the flux ratio \ion{C}{4}/(Ly$\alpha$ +
\ion{N}{5}) was very low for Q0353-383, only 0.07 compared to average
values of $\sim 0.30$, indicating a low carbon abundance. For the seven
objects in this work with $z > 2.3$, the ratio of \ion{C}{4}/(Ly$\alpha$
+ \ion{N}{5}) ranges from 0.09 to 0.20, with a median value of 0.17.
The typical error on the measurement of this ratio for each object is
only $\sim 0.02$, so it seems that the nitrogen-rich candidates
presented here have slightly larger relative carbon abundances than
Q0353-383, while still being slightly depressed relative to the rest of
the quasar population, as demonstrated by the somewhat higher value of
0.31 measured for the SDSS quasar composite.  It is interesting to note
that the object with the lowest value of this ratio (0.09) is SDSS
J1550+0236, which appears to have much stronger \ion{N}{4}] and
\ion{N}{3}] lines than the other six objects for which this ratio was
calculated (see Table 2).

A composite spectrum was generated using all twenty of the candidate
nitrogen-rich quasars.  Each spectrum was divided by a fit to the
continuum, scaling them all to a common continuum level of one.  The
scaled spectra were then averaged together, weighted by the S/N in the
original spectra.  Because of the various redshifts of the quasars, the
region of overlap for all of the spectra is from $\sim 1450 - 2250$
\AA.  Figure 3 displays the resulting composite overlaid by the SDSS
composite from the EDR generated by \citet{VandenBerk2001}, which has
also been scaled by dividing out the continuum.  This particular method
is useful for comparing emission lines individually, but obviously
erases any information about the differences in the shapes of the
continua of the two composites.  The nitrogen-rich quasars are much
stronger in the \ion{N}{4}] $\lambda$1486 and \ion{N}{3}] $\lambda$1750
emission lines, as well as Ly$\alpha$.  \ion{C}{4} and \ion{C}{3}] seem
to be slightly stronger in the nitrogen-rich quasars as well, although
the FWHM of the \ion{C}{4} line is much smaller in the nitrogen-rich
composite than in the SDSS composite, with values of $\sim 2560$ km
s$^{-1}$ and $\sim 4850$ km s$^{-1}$, respectively.  

Based on the equivalent widths of their \ion{N}{4}] $\lambda$1486 and
\ion{N}{3}] $\lambda$1750 emission lines, four of the 20 candidate
nitrogen-rich quasars may be viewed as being similar to Q0353-383: SDSS
J0909+5803, SDSS J1254+0241, SDSS J1546+5253, and SDSS J1550+0236.
Figure 4 compares these four quasars, which all have $W_{N III} \gtrsim
7.0$\AA\ and $W_{N IV} \gtrsim 4.0$\AA, compared to $W_{N III} =
9.0$\AA\ and $W_{N IV} = 5.0$\AA\ for Q0353-383.

As indicated in Table 1, several of the nitrogen-rich candidates
discussed in this paper are also included in the SDSS EDR Quasar Catalog
\citep{Schneider2002}, although the four candidates mentioned above with
the strongest emission were discovered in the area of the DR1 not
coincident with the EDR. For most of the objects that were not detected
by \citet{Bentz2003} in their study of quasars in the EDR, the EDR
spectra show obvious emission from \ion{N}{3}] $\lambda$1750.  However,
there is only weak evidence for emission from \ion{N}{4} $\lambda 1486$,
and this is easily accounted for by the noise in the spectra.
Fortunately, the SDSS spectroscopic pipeline has gone through several
changes resulting in much cleaner spectra in the DR1 database, as shown
by the comparison of the final EDR and DR1 spectra for SDSS J1136+0110
in Figure 5.  What originally appeared to be noise in the EDR spectra of
these objects is now more definitely shown to be weak emission.
Additionally, SDSS J1130-0045 was not detected by \citet{Bentz2003}
because of the slight difference in the redshift cuts between the two
searches.  In this work, we have examined quasars with $1.6 < z < 4.1$,
while \citet{Bentz2003} only examined quasars with $1.8 < z < 4.1$.  At
a redshift of 1.66, SDSS J1130-0045 would not have been included in the
EDR study.

\section{Conclusions}
We have searched 6650 quasars in the SDSS DR1 database for nitrogen-rich
objects similar to Q0353-383.  Four candidates have nitrogen emission as
strong or stronger than that seen in Q0353-383, and an additional 16
exhibit slightly weaker nitrogen emission. We have also identified 33
objects that may have visible \ion{N}{4}] and \ion{N}{3}] emission,
although it is less clear from the quality of their spectra.

With the data available, we may set a lower limit of 0.06\% (about 1 in
1700) on the number of nitrogen-rich objects that are truly similar to
Q0353-383.  If we view Q0353-383 and its companion SDSS quasars as being
the most extreme objects in a continuous phase of nitrogen enrichment,
then it also appears that a lower limit of 0.2\% - 0.7\% of quasars (up
to 1 in 130) are approaching the nitrogen enrichment levels of their
more extreme counterparts.

If nitrogen-strong quasars are quasars viewed at the peak of metal
enrichment, then the length of that phase is approximately 1/1700th of
the typical quasar lifetime.  For example, for a quasar lifetime of
10$^7$ years, these objects would be viewed only in the last 6,000 years
of their existence as quasars.  Alternatively, it may be that only 1 in
1700 quasars reaches the extremely high metallicities needed to generate
strong nitrogen emission.

Further data and modeling outside the scope of this paper are needed to
place these quasars in their correct context to the overall quasar
population.  For example, if these quasars are near the end of their
accretion phases, or are the most metal-rich because they are in the
most massive bulges, their black holes should be biased towards higher
masses than randomly selected quasars.  However, there is no evidence of
such a bias in the widths of the emission lines, which range from narrow
to average.

Additional observations should be undertaken in order to obtain higher
S/N spectra for more accurate measurements of the emission lines, and
also to push the observed wavelengths further into the blue in order to
gain the Ly$\alpha$ and \ion{N}{5} emission lines for those objects with
$z \approx 2$.  With such additional data, metallicity estimates may be
made using line ratios such as \ion{N}{3}]/\ion{O}{3}],
\ion{N}{3}]/\ion{C}{3}], and \ion{N}{5}/\ion{He}{2}.  These estimates
would allow us to test the hypothesis that nitrogen-rich quasars are
exhausting their fuel supplies and approaching the metallicities
predicted by numerical simulations for black holes as they end their
quasar activity.

\acknowledgements
We would like to thank Joe Shields for helpful comments and
conversations. We are also very grateful to the referee, Don Schneider,
for suggestions that considerably improved the presentation and value of
this paper.

Funding for the creation and distribution of the SDSS Archive has been
provided by the Alfred P. Sloan Foundation, the Participating
Institutions, the National Aeronautics and Space Administration, the
National Science Foundation, the U.S. Department of Energy, the Japanese
Monbukagakusho, and the Max Planck Society. The SDSS Web site is
\url{http://www.sdss.org/}.

The SDSS is managed by the Astrophysical Research Consortium (ARC) for
the Participating Institutions. The Participating Institutions are The
University of Chicago, Fermilab, the Institute for Advanced Study, the
Japan Participation Group, The Johns Hopkins University, Los Alamos
National Laboratory, the Max-Planck-Institute for Astronomy (MPIA), the
Max-Planck-Institute for Astrophysics (MPA), New Mexico State
University, University of Pittsburgh, Princeton University, the United
States Naval Observatory, and the University of Washington.

\bibliographystyle{apj}
\bibliography{MistyBentz}

\clearpage

\begin{figure}
\figurenum{1}
\epsscale{0.5}
\plotone{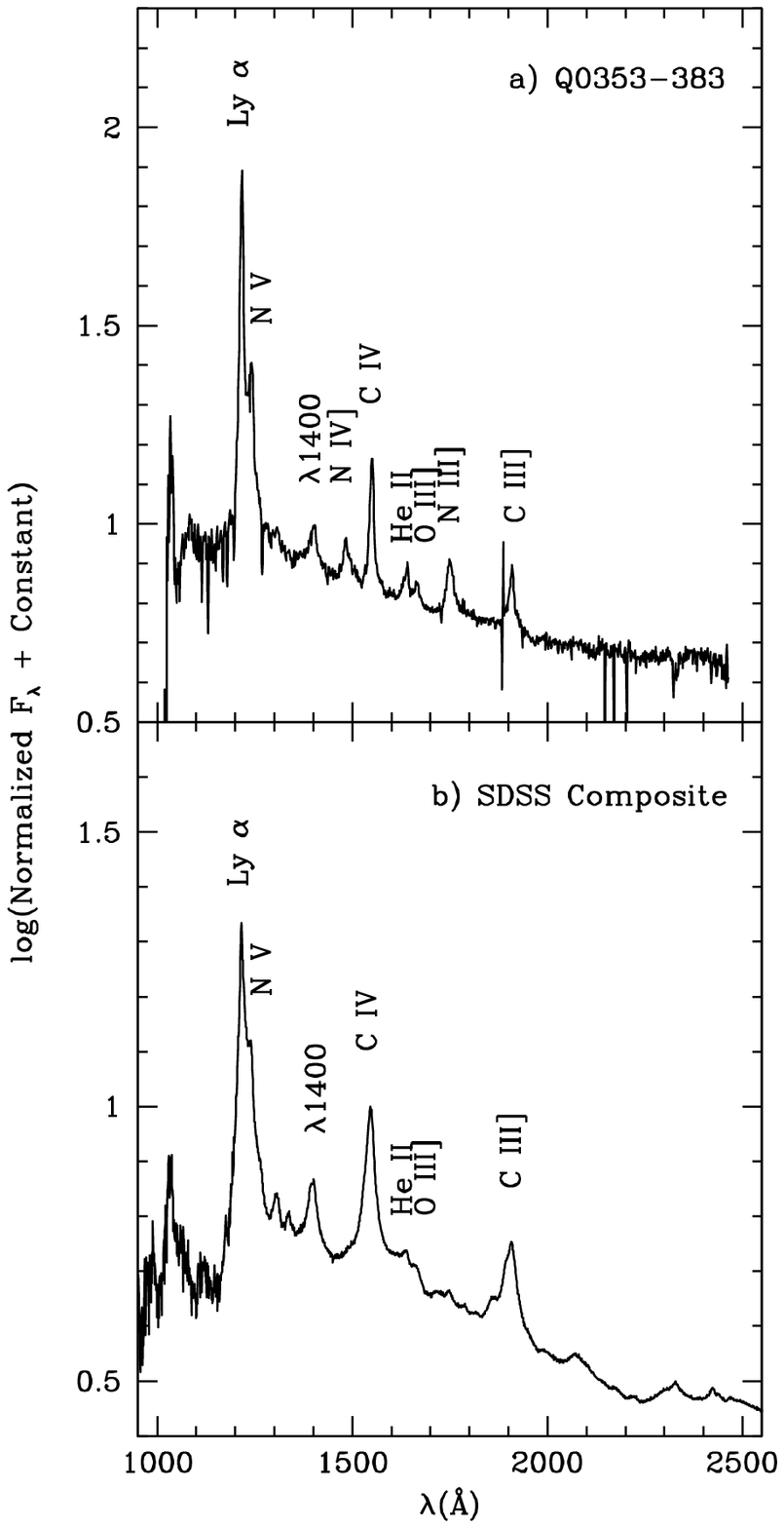}
\caption{Rest frame spectra of a.) Q0353-383 (Baldwin, private 
         communication), and b.) the SDSS composite
         \citep{VandenBerk2001}, composed of 2204 quasar spectra.  Both
         spectra are plotted in semi-log format to enhance fine
         details.}
\epsscale{1.0}
\end{figure}

\begin{figure}
\figurenum{2}
\plotone{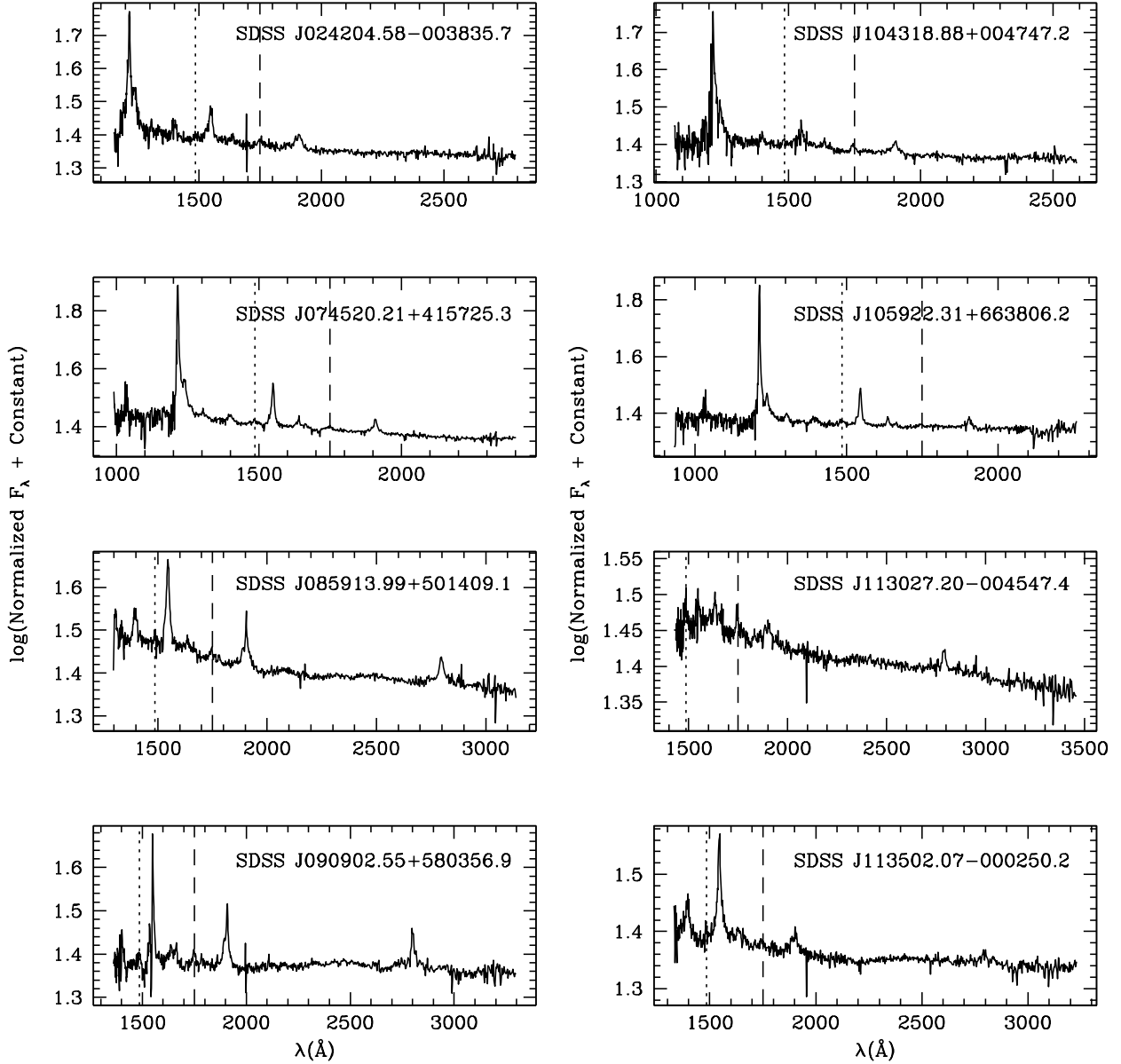}
\caption{Rest frame spectra of the 20 candidate nitrogen-rich quasars 
         from the SDSS DR1 database.  The spectra are smoothed with a
         bin of five pixels and plotted in semi-log format to enhance
         fine details.  The dotted line is at \ion{N}{4}] $\lambda$1486
         and the dashed line is at \ion{N}{3}] $\lambda$1750.  Several
         of the spectra retain the signatures of night sky lines, but a
         handful of them also seem to have narrow absorption profiles.
         SDSS J0909+5803 and SDSS J1254+0241 have blueshifted absorption
         in \ion{Si}{4}, \ion{C}{4}, and \ion{Mg}{2}.  SDSS J2040-0545
         has blueshifted absorption in \ion{Si}{4}, \ion{C}{4}, and
         \ion{Al}{3}, and SDSS J1327+0035 has two blueshifted \ion{C}{4}
         absorption systems.  SDSS J1043+0047 has an associated
         absorption system in Ly$\alpha$, \ion{N}{5}, \ion{Si}{4}, and
         \ion{C}{4}, as well as two other associated \ion{C}{4}
         absorbers.  And SDSS J1546+5253 has an intervening \ion{Mg}{2}
         and \ion{Fe}{2} system at $z=0.792$, producing absorption in
         the \ion{Si}{4} and \ion{C}{4} emission lines.}
\end{figure}
\begin{figure}
\figurenum{2}
\plotone{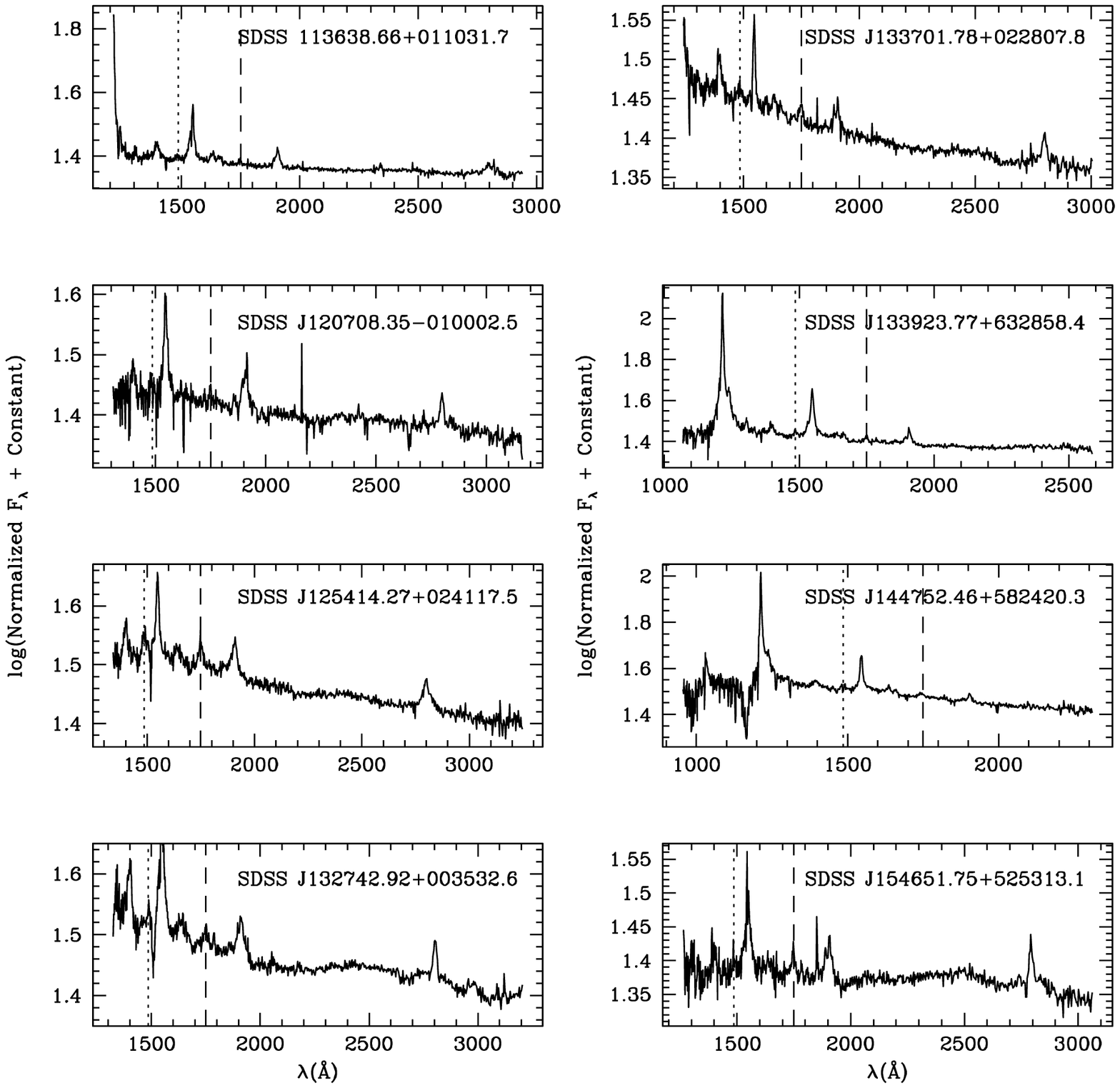}
\caption{\it{continued}}
\end{figure}
\begin{figure}
\figurenum{2}
\plotone{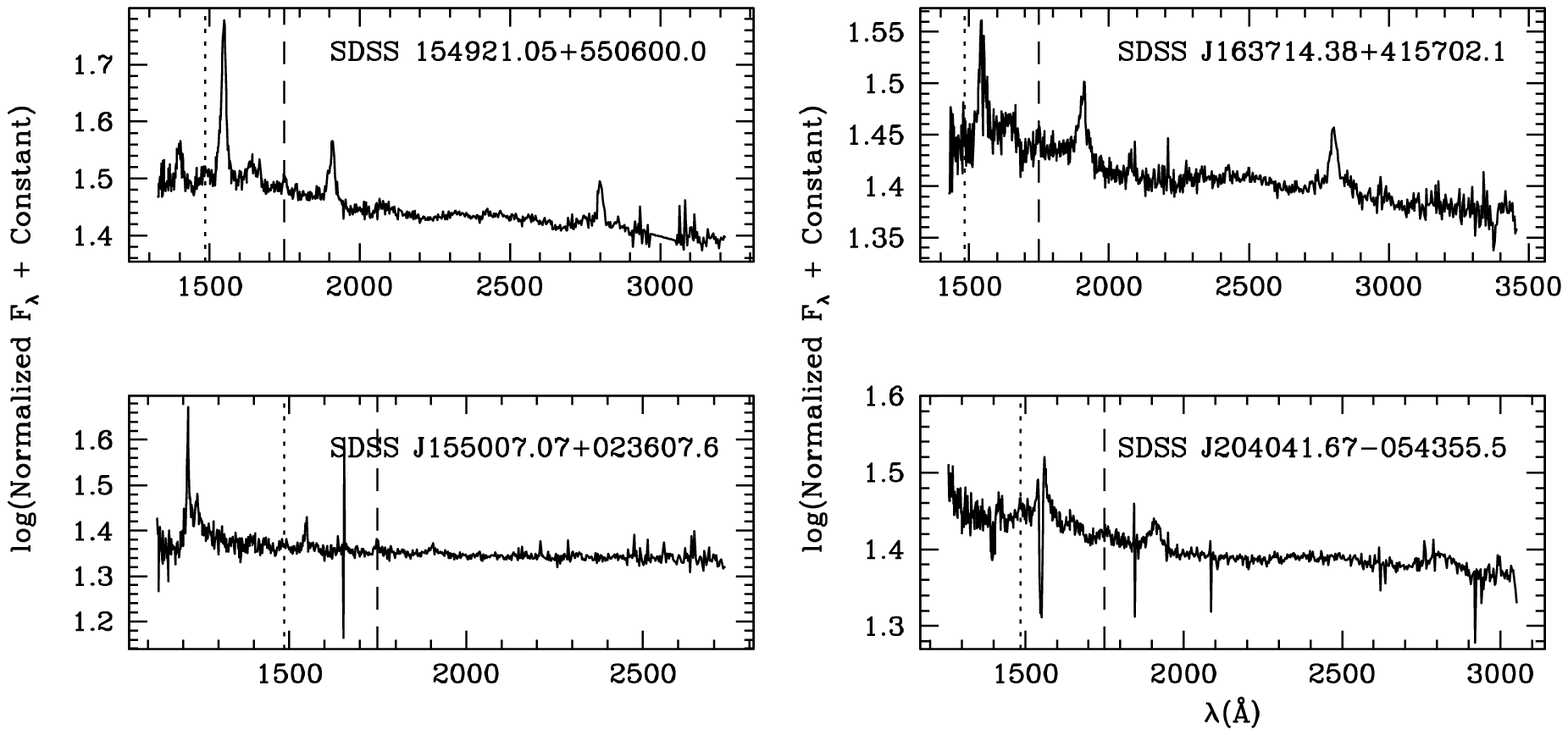}
\caption{\it{continued}}
\end{figure}

\begin{figure}
\figurenum{3}
\plotone{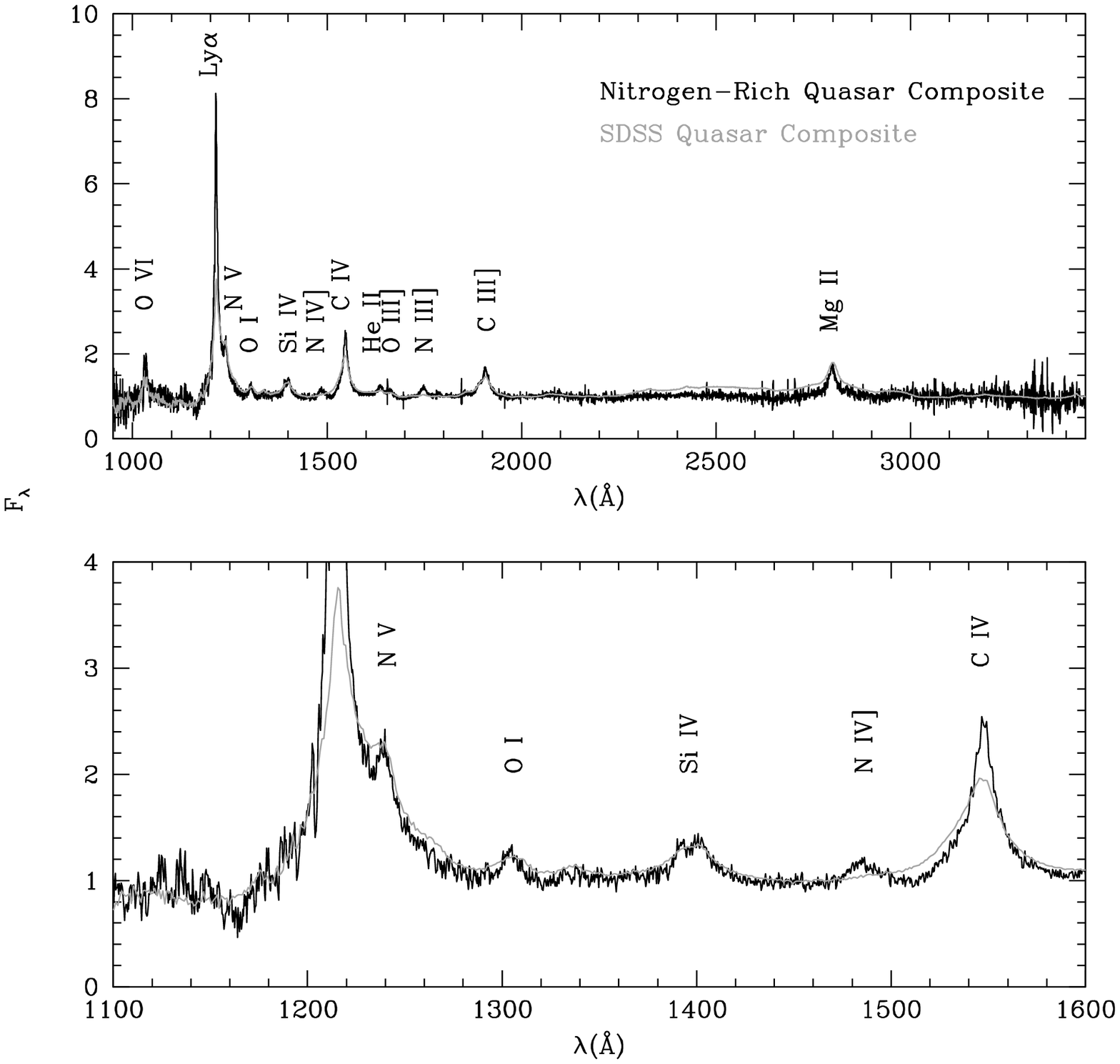}
\caption{Comparison of the candidate nitrogen-rich quasar composite 
         (black) and the SDSS quasar composite (gray).  Both spectra are
         scaled to a continuum level of 1.  The bottom window is simply
         a rescaled view of the region between 1100\AA\ and 1600\AA.}
\end{figure}

\begin{figure}
\figurenum{4}
\plotone{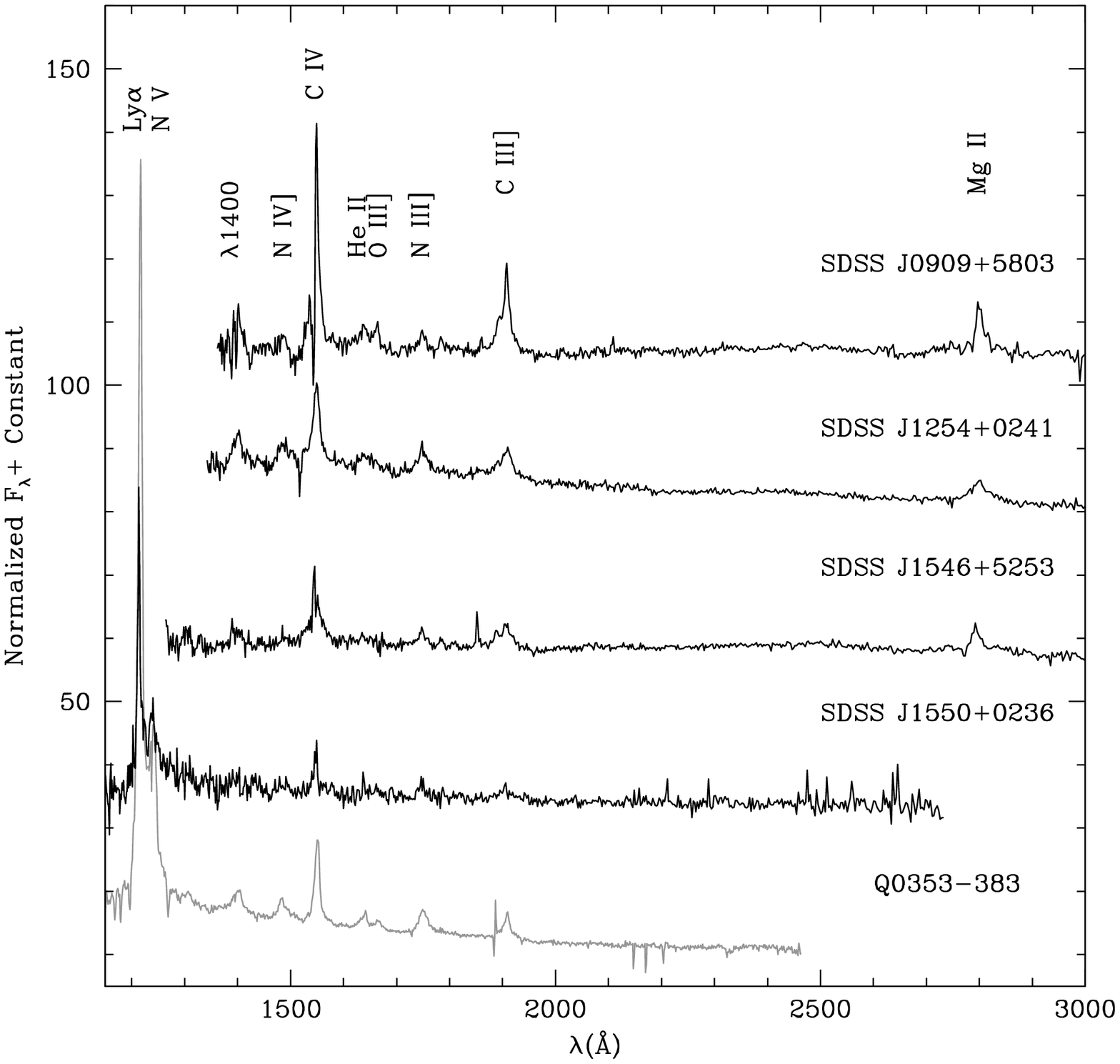}
\caption{Comparison of Q0353-383 (plotted in gray) with the four 
	 strongest candidate nitrogen-rich quasars discovered in the
	 DR1.  The SDSS spectra have been smoothed with a bin of five
	 pixels, and egregious night sky lines have been removed.}
\end{figure}

\begin{figure}
\figurenum{5}
\plotone{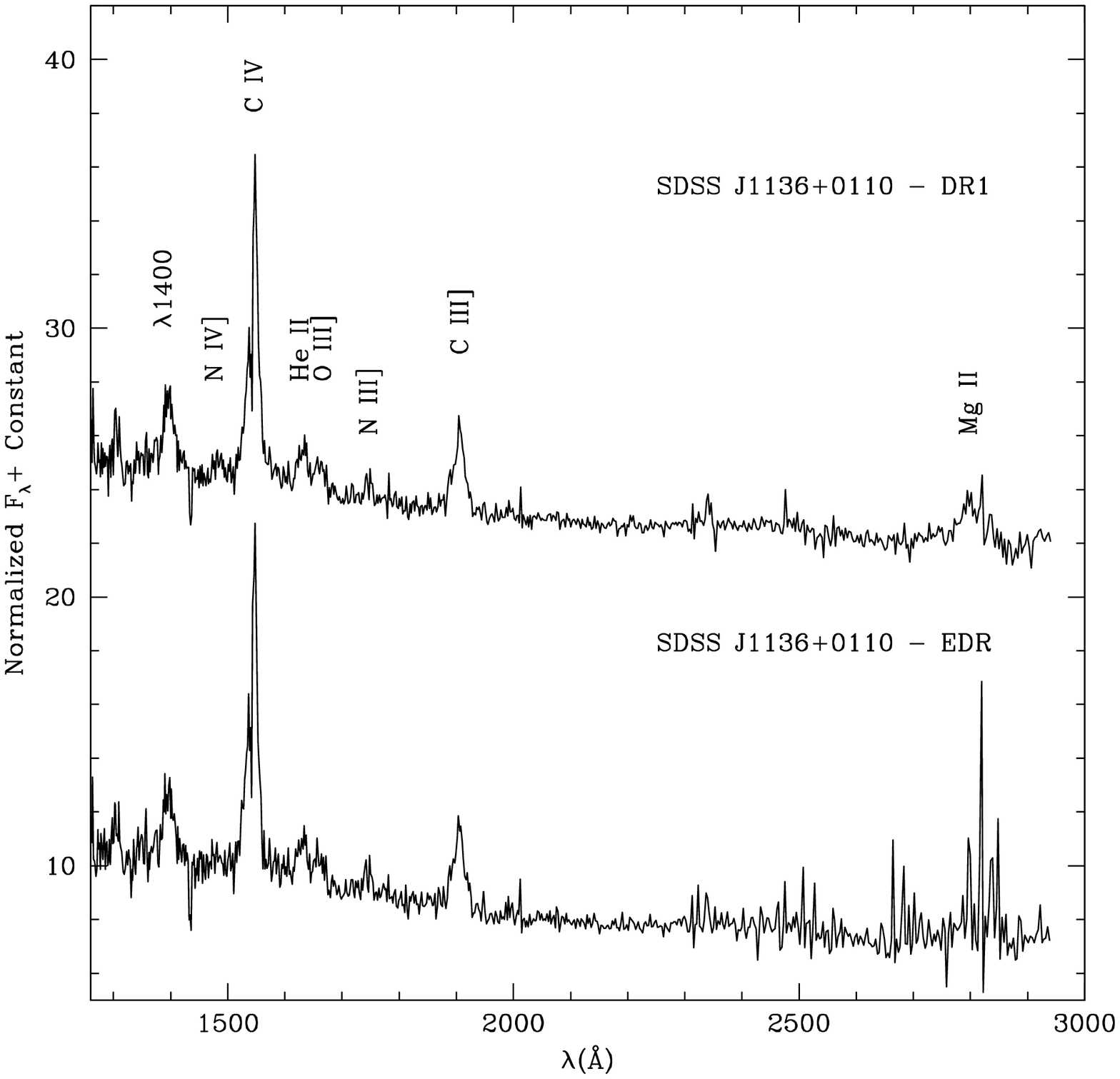}
\caption{Comparison of the EDR (bottom) and DR1 (top) spectra of the
         candidate nitrogen-rich quasar SDSS J1136+0110.  Each spectrum 
         has been smoothed with a bin of five pixels.}
\end{figure}

\clearpage

\begin{deluxetable}{lccclcl} 
\tabletypesize{\footnotesize}
\tablecolumns{7} 
\tablewidth{0pc}
\tablecaption{Properties of Candidate Nitrogen-Rich Quasars}
\tablehead{ 
\colhead{Quasar (SDSS J)} & 
\colhead{z\tablenotemark{a}} & 
\colhead{i\tablenotemark{a}} & 
\colhead{M$_i$\tablenotemark{b}} & 
\colhead{Additional Identifier} & 
\colhead{Notes\tablenotemark{c}} &
\colhead{References\tablenotemark{d}}} 
\startdata
024204.58$-$003835.7 	& 2.288 & 19.855 & -25.88 & 
	SDSS J024204.59$-$003835.7 & EDR & 1\\ 
074520.21+415725.3 	& 2.843 & 19.116 & -27.14 & & FIRST & 2 \\ 
085913.99+501409.1 	& 1.930 & 18.932 & -26.40 & & & \\ 
090902.55+580356.9 	& 1.793 & 18.986 & -26.21 & & & \\ 
104318.88+004747.2 	& 2.558 & 19.277 & -26.74 & & EDR & 1\\ 
105922.31+663806.2	& 3.075	& 19.874 & -26.49 & & & \\
113027.20$-$004547.4 	& 1.662 & 18.752 & -26.24 & 
	2QZ 113027.2$-$004548 & EDR, NRD & 1, 3, 4\\ 
113502.07$-$000250.2 	& 1.851 & 19.599 & -25.63 & 
	SDSS J113502.08$-$000250.1 & EDR & 1\\ 
113638.66+011031.7 	& 2.131 & 19.449 & -26.10 & & EDR & 1\\ 
120708.35$-$010002.5 	& 1.912 & 18.627 & -26.68 & & EDR & 1\\ 
125414.27+024117.5 	& 1.839 & 18.319 & -26.92 & & FIRST & 2 \\ 
132742.92+003532.6 	& 1.873 & 18.333 & -26.93 & & EDR & 1, 5\\ 
133701.78+022807.8 	& 2.068 & 19.167 & -26.32 & & FIRST & 2 \\ 
133923.77+632858.4 	& 2.559 & 19.270 & -26.69 & 
	87GB 133745.0+634401 & FIRST, FSRQ, SLC & 2, 4, 6, 7 \\ 
144752.46+582420.3 	& 2.982 & 18.214 & -28.07 & & FIRST, 2MASS & 2, 8 \\ 
154651.75+525313.1 	& 2.011 & 19.057 & -26.35 & & & \\ 
154921.05+550600.0 	& 1.865 & 18.422 & -26.81 & & FIRST & 2 \\ 
155007.07+023607.6 	& 2.371 & 20.064 & -25.94 & & FIRST & 2 \\ 
163714.38+415702.1 	& 1.665 & 18.784 & -26.17 & & & \\ 
204041.67$-$054355.5 	& 2.023 & 18.901 & -26.61 & & & \\
\enddata
\tablenotetext{a} {Values taken from \citet{Schneider2003}}
\tablenotetext{b} {As determined by \citet{Schneider2003}, with 
                   $H_{0} = 70$, $\Omega_M = 0.3$, $\Omega_{\Lambda} =
                   0.7$, and $\alpha_Q = -0.5$ }
\tablenotetext{c} {\bf{Abbreviations}: \rm EDR = SDSS Early Data Release 
		source; FIRST = Faint Images of the Radio Sky at
		Twenty-Centimeters source; NRD = not detected in radio;
		FSRQ = flat spectrum radio quasar; SLC = superluminal
		candidate; 2MASS = Two Micron All Sky Survey source}
\tablenotetext{d} {This research has made use of the NASA/IPAC 
                   Extragalactic Database (NED) which is operated by the
                   Jet Propulsion Laboratory, California Institute of
                   Technology, under contract with the National
                   Aeronautics and Space Administration.}
\tablerefs{1. \citet{Schneider2003}, 
           2. \citet{Becker2003},
	   3. \citet{Croom2001},
           4. \citet{Veron-Cetty2001},
           5. \citet{Bentz2003},
	   6. \citet{Vermeulen1995}, 
           7. \citet{Becker1991},
	   8. \citet{Cutri2003}}.

\end{deluxetable}

\begin{deluxetable}{lccccccc}
\tabletypesize{\footnotesize}
\tablecolumns{8} 
\tablewidth{0pc}
\tablecaption{Measured Equivalent Widths \ion{N}{4}] $\lambda$1486 and \ion{N}{3}] $\lambda$1750 }
\tablehead{
\colhead{}  & 
\colhead{}  & 
\multicolumn{3}{c}{\ion{N}{4}]} & 
\multicolumn{3}{c}{\ion{N}{3}]} \\ 
\cline{3-8} \\
\colhead{Quasar} & 
\colhead{S/N\tablenotemark{a}} & 
\colhead{$W$(\AA)\tablenotemark{b}} & 
\colhead{$\sigma$\tablenotemark{c}} & 
\colhead{CCC\tablenotemark{d}} &
\colhead{$W$(\AA)\tablenotemark{b}} & 
\colhead{$\sigma$\tablenotemark{c}} &
\colhead{CCC\tablenotemark{d}}}
\startdata
Q0353$-$383\tablenotemark{e}	& \nodata & 5.0 & \nodata & 0.98 & 9.0 & \nodata & 0.99\\
SDSS J0242$-$0038 & 5.0  & 2.0 & 1.8  & 0.22  & 8.0 & 7.3  & 0.68 \\ 
SDSS J0745+4157   & 8.1  & 3.0 & 4.4  & 0.39 & 4.0  & 5.9  & 0.86 \\
SDSS J0859+5014	  & 9.4  & 2.0 & 3.4  & 0.35 & 3.0  & 5.1  & 0.76 \\
SDSS J0909+5803   & 4.1  & 4.0 & 3.0  & 0.31 & 8.0  & 6.0  & 0.85 \\
SDSS J1043+0047   & 8.4  & 3.0 & 4.6  & 0.41 & 5.0  & 7.7  & 0.71 \\
SDSS J1059+6638	  & 5.6  & 4.0 & 4.1  & 0.47 & 4.0  & 4.1  & 0.59 \\	
SDSS J1130$-$0045 & 8.5  & 3.0 & 4.7  & 0.00 & 4.0  & 6.2  & 0.75 \\
SDSS J1135$-$0002 & 5.0  & 4.0 & 3.7  & 0.06 & 6.0  & 5.5  & 0.80 \\
SDSS J1136+0110   & 6.8  & 3.0 & 3.7  & 0.22 & 3.0  & 3.7  & 0.53 \\
SDSS J1207$-$0100 & 5.7  & 5.0 & 5.2  & 0.31 & 3.0  & 3.1  & 0.66 \\
SDSS J1254+0241   & 11.3 & 7.0 & 14.4 & 0.68 & 8.0  & 16.5 & 0.93 \\
SDSS J1327+0035   & 9.6  & 3.0 & 5.3  & 0.46 & 5.0  & 8.8  & 0.94 \\
SDSS J1337+0228   & 11.8 & 3.0 & 6.5  & 0.52 & 4.0  & 8.6  & 0.83 \\
SDSS J1339+6328   & 8.6  & 4.0 & 6.3  & 0.63 & 4.0  & 6.3  & 0.70 \\
SDSS J1447+5824   & 16.1 & 2.0 & 5.9  & 0.12 & 3.0  & 8.8  & 0.78 \\
SDSS J1546+5253   & 4.6  & 4.0 & 3.4  & 0.43 & 11.0 & 9.2  & 0.91 \\
SDSS J1549+5506   & 10.5 & 5.0 & 9.6  & 0.35 & 2.0  & 3.8  & 0.84 \\
SDSS J1550+0236   & 3.5  & 7.0 & 4.5  & 0.36 & 7.0  & 4.5  & 0.41 \\
SDSS J1637+4157   & 5.6  & 2.0 & 2.0  & 0.00 & 3.0  & 3.1  & 0.63 \\
SDSS J2040$-$0545 & 7.7  & 3.0 & 4.2  & 0.71 & 6.0  & 8.4  & 0.65 \\
\enddata
\tablenotetext{a} {Signal-to-noise ratio of the SDSS spectrum, measured 
	           between $\lambda1675$ and $\lambda1725$\AA\ before
	           smoothing.}
\tablenotetext{b} {Guideline measurements in the quasar rest frame.  
                   Errors are estimated to be 0.5-1.0 \AA.}
\tablenotetext{c} {Significance of detection, assuming a 30\AA\ window in 
		   the quasar rest frame}
\tablenotetext{d} {Cross-correlation coefficients for the 30\AA\ 
                   windows as described in the text; auto-correlation
                   coefficients for Q0353-383}
\tablenotetext{e} {Values taken from \citet{Baldwin2003}}  
\end{deluxetable}

\begin{deluxetable}{lccclcl} 
\tabletypesize{\footnotesize}
\tablecolumns{7} 
\tablewidth{0pc}
\tablecaption{Properties of Nitrogen-Salient Candidate Quasars}
\tablehead{ 
\colhead{Quasar (SDSS J)} & 
\colhead{z\tablenotemark{a}} & 
\colhead{i\tablenotemark{a}} & 
\colhead{M$_i$\tablenotemark{b}} & 
\colhead{Additional Identifier} & 
\colhead{Notes\tablenotemark{c}} &
\colhead{References\tablenotemark{d}}} 
\startdata
003815.93+140304.6	& 2.718	& 19.794 & -26.39 & & & \\
004833.56+142056.8 	& 1.816 & 18.608 & -26.67 & & & \\ 
010527.72+143701.3 	& 1.939 & 19.171 & -26.22 & & & \\ 
012310.96$-$084053.7 	& 1.634 & 19.271 & -25.71 & & & \\ 
014517.79+135602.2 	& 1.949 & 18.866 & -26.55 & & & \\ 
022129.05$-$083351.0	& 1.876 & 20.170 & -25.10 & & & \\
022203.24$-$091531.4 	& 1.742 & 18.948 & -26.14 & & & \\ 
025713.07$-$010157.8	& 1.869 & 19.190 & -26.15 & 
	[CLB91] 025440.2-011 & EDR, pROSAT & 1, 2, 3, 4 \\
025939.26$-$063158.4 	& 2.012 & 19.471 & -26.04 & & & \\ 
033832.65+004518.5 	& 1.839 & 19.240 & -26.10 & 
	SDSS J033832.65+004518.4 & EDR & 1 \\
040913.78+060839.0	& 2.685 & 19.210 & -26.98 & & & \\
073048.37+371616.3	& 1.676 & 19.672 & -25.39 & & FIRST & 5 \\
073506.59+374014.1	& 2.012 & 18.966 & -26.52 & & & \\
091031.35+010151.9 	& 2.013 & 18.676 & -26.76 & & & \\
091745.24+555935.0 	& 1.961 & 19.060 & -26.31 & & & \\ 
092225.69+611530.6	& 2.227 & 18.472 & -27.19 & & & \\
092541.46+004040.9 	& 1.935 & 19.681 & -25.67 & & & \\ 
092754.17+582811.6	& 1.847 & 18.775 & -26.47 & & & \\
093510.60+593937.3 	& 1.816 & 19.142 & -26.04 & & & \\ 
102949.50+643835.9 	& 1.870 & 19.781 & -25.45 & & & \\ 
103013.61+010056.4	& 2.053 & 18.882 & -26.65 & & EDR & 1 \\
113012.38+003314.7 	& 1.932 & 19.492 & -25.85 & & EDR & 1\\ 
115357.27+002754.0	& 1.672 & 19.094 & -25.90 & & EDR & 1\\
121933.26+003226.4 	& 2.871 & 19.341 & -26.89 & 
	SDSS J121933.25+003226.5 & EDR, FIRST & 1, 5\\ 
122348.22+010221.9	& 1.942 & 18.616 & -26.74 & & EDR & 1 \\
124727.25+033009.9	& 1.635 & 19.120 & -25.82 & & & \\
125134.59+681824.2 	& 1.662 & 19.055 & -25.91 & & & \\ 
125925.03+642139.6 	& 1.948 & 19.235 & -26.10 & & & \\ 
141857.12+635524.4	& 3.117 & 19.532 & -26.86 & & & \\
155003.71+031325.0 	& 1.788 & 17.648 & -27.75 & & FIRST, 2MASS & 5, 6\\ 
210255.16+064112.2	& 2.109 & 19.289 & -26.38 & & & \\	
\enddata
\tablenotetext{a} {Values taken from \citet{Schneider2003}}
\tablenotetext{b} {As determined by \citet{Schneider2003}, with 
                   $H_{0} = 70$, $\Omega_M = 0.3$, $\Omega_{\Lambda} =
                   0.7$, and $\alpha_Q = -0.5$}
\tablenotetext{c} {\bf Abbreviations: \rm EDR = SDSS Early Data Release 
		   source; pROSAT = pointed ROentgen SATellite source;
		   FIRST = Faint Images of the Radio Sky at
		   Twenty-Centimeters source; 2MASS = Two Micron All Sky
		   Survey source}
\tablenotetext{d} {This research has made use of the NASA/IPAC 
                   Extragalactic Database (NED) which is operated by the
                   Jet Propulsion Laboratory, California Institute of
                   Technology, under contract with the National
                   Aeronautics and Space Administration.}
\tablerefs{ 1. \citet{Schneider2003},
	    2. \citet{Vignali2003},
	    3. \citet{Veron-Cetty2001},
	    4. \citet{Cristiani1989},
	    5. \citet{Becker2003},
	    6. \citet{Cutri2003}}.
\end{deluxetable}

\end{document}